\documentclass[%
floatfix,
reprint,
superscriptaddress,
 amsmath,amssymb,
 aps,
pre,
]{revtex4-2}

\usepackage{graphicx}% Include figure files
\usepackage{dcolumn}% Align table columns on decimal point
\usepackage{bm}% bold math
\usepackage{mathtools}
\usepackage{amssymb, amsmath}
\usepackage{dsfont}
\usepackage{physics2}
\usephysicsmodule{ab}
\usepackage{derivative}
\usepackage{mathrsfs}
\usepackage{enumitem}
\usepackage{xcolor}
\usepackage{xfrac}

\newcommand{\indv}[2][]{\text{d}#1/\text{d}#2}
\begin{document}

\title{Collisional stopping power of ions in warm dense matter}% Force line breaks with \\

\author{Lucas Babati}
\affiliation{Nuclear Engineering and Radiological Sciences, University of Michigan, Ann Arbor, Michigan, 48109, USA}%

\author{Shane Rightley}%
\affiliation{Sandia National Laboratory, Albuquerque, New Mexico, 87123, USA}

\author{Nathaniel Shaffer}
\affiliation{Laboratory for Laser Energetics, University of Rochester, Rochester, New York, 14623, USA}%

\author{Scott Baalrud}
\affiliation{Nuclear Engineering and Radiological Sciences, University of Michigan, Ann Arbor, Michigan, 48109, USA}%
\email{baalrud@umich.edu}

\date{\today}% It is always \today, today,

%Ion stopping in matter is an important process in many physical situations,
%especially for inertial confinement fusion (ICF) where alpha particles 
%stop on the surrounding plasma, transferring energy to the plasma and 
%maintaining the reaction. Thus it is important to study how charges particle
%stop in materials relevant to ICF. One phase of increased study recently has 
%been Warm Dense Matter (warm dense matter), an exotic state of matter with strongly correlated 
%ions and quantum mechanical electrons. We present a kinetic theory based approach
%to calculate stopping power in warm dense matter that captures the correlation physics and 
%necessary quantum physics. This method is shown to be comparable in accuracy 
%to solid state theories such as Density Functional Theory Molecular Dynamics (DFTMD),
%while being much less computationally expensive to calculate. We then examine 
%the microscopic physics and show that our theory captures the transition from 
%a plasma state where ballistic ions stop on thermal electrons, to the degenerate
%regime where the ions instead stop on electrons around the Fermi energy.
\begin{abstract}
    A model for the collisional stopping of ions on free electrons in warm dense matter is developed and explored. 
    It is based on plasma kinetic theory, but with modifications to address the warm dense matter regime. 
Specifically, it uses the Boltzmann-Uehling-Uhlenbeck kinetic equation to incorporate effects of Fermi degeneracy of electrons. The cross section is computed from quantum scattering of electrons and ions occuring via the potential of mean force derived from an average atom model, which incorporates effects of strong Coulomb correlations. 
%to push plasma theory into the strongly coupled and degenerate regime where warm dense matter exists.
Predictions from this model show comparable accuracy to results from time-dependent density functional theory calculations for deuterium near solid density and a temperature of several electronvolts, at a fraction of the computational cost. 
Further, the model 
    captures the transition of a plasma from the classical limit to the degenerate
    limit, including qualitative behaviors of solid state theory. 
    % This transition from classical to degenerate behavior is associated with the characteristic relative energy of scattering particles transitioning from 
    % the classical thermal energy to the Fermi energy. 
    %This transition is quantified through changes
    %in the electron-ion collision frequency in the two limits. 
\end{abstract}

\maketitle

\section{Introduction}
The stopping of ions in various materials is a common and well understood problem 
in many areas of physics~\cite{Bethe1930, Srim}. In plasma physics, the study of stopping power
has shown increased importance in the effort for fusion energy using inertial-confinement
fusion (ICF) \cite{BettiNature2016}. In ICF, energetic charged particles must deposit the energy gained from 
a fusion reaction, or from a particle beam, into the burning plasma to promote more reactions
 \cite{ZylstraPop2019, KeyLLNL2005}. This energy transfer is almost entirely due to a drag
force the charged particles experience from the surrounding plasma. Accordingly,
accurately modeling stopping power in materials relevant to ICF is important for predicting the behavior of these experiments. Outside of the ICF application, 
stopping power is a quantity which can be directly measured by experiment 
\cite{ZylstraPRL2015, CayzacPRE2015, Cayzacnaturecomm2017, FrenjePRL2019, HayesNature2020, MalkoNature2022, LahmannPPCF2023}. Since it is a result of ion-electron collisions, it is also a proxy for 
other transport properties determined by electron-ion collisions such as electrical and thermal conductivities. 
Thus, it is a good example property to test transport theory. 
%Thus, it is important to relate calculations of stopping power to these microscopic 
%collision rates as a means to test transport theories across plasma physics.

Here, a model is presented to calculate stopping power in the warm dense matter
regime. The method applies the Boltzmann-Uehling-Uhlenbeck (BUU) \cite{UehlingPRL1933, UehlingPRL1934} equation, but
treats ion-electron interactions with the potential of mean force rather than the Coulomb potential or screened Coulomb (Debye) potential~\cite{Hansen, StarrettPRE2013, StarrettHEDP2017}. 
With this combination, the degeneracy and strong correlation effects relevant to the warm dense matter regime are self-consistently 
accounted for. The resulting model may be viewed in two ways: as an extension of the 
theory of weakly coupled degenerate gases to the strongly coupled regime with 
the use of the potential of mean force, or as an extension of mean force kinetic theory~\cite{BaalrudPRL2013, BaalrudPOP2019}
into the degenerate regime by generalizing the kinetic equation from a Boltzmann
operator to the BUU operator. Further, the approach introduced
in Ref.~\cite{RightleyPRE2021} has been generalized to arbitrary velocities for the interacting particles, enabling the calculation of stopping power.

\begin{figure}
    \includegraphics[width=0.48\textwidth]{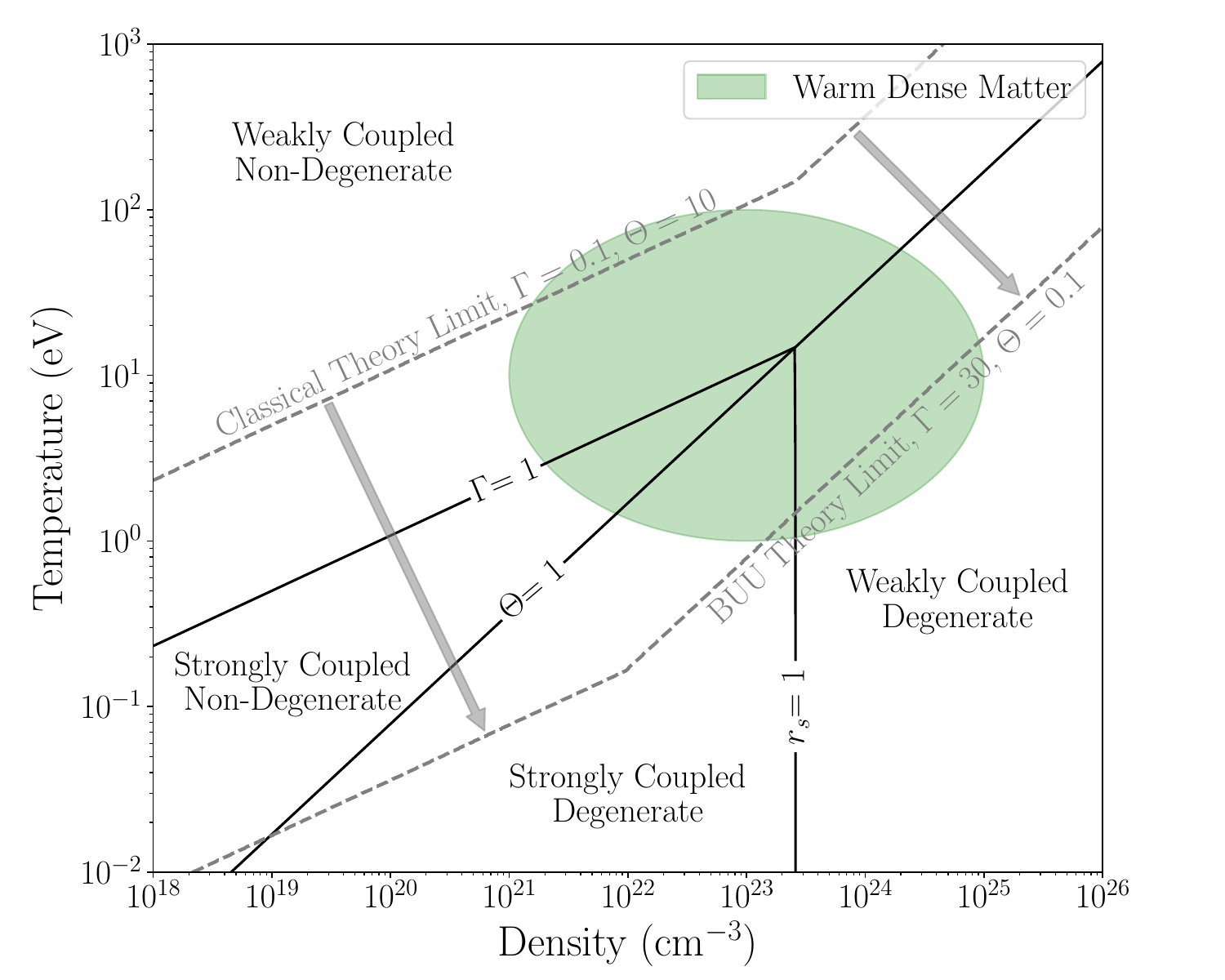}
    \caption{\label{fig:phase_space} The density temperature phase space around
        the warm dense matter regime denoted by the green shaded region. The 
        phase space is split up by the lines $\Gamma=1$, $\Theta=1$, and $r_s =1$, into 
        four regions determined by the dominate energy in the system.
        The two dashed gray lines mark where classical plasma physics theories 
        fail at $\Gamma=0.1$ and $\Theta=10$ and where the Boltzmann-Uehling-Uhlenbeck
    based theory extends this to at $\Gamma=30$ and $\Theta=0.1$.}
\end{figure}

Warm dense matter is typically defined by a region of density and temperature parameter space where
three dimensionless parameters merge; see Figure~\ref{fig:phase_space}.
The first is the Coulomb coupling parameter, defined by 
\begin{equation}
    \Gamma_{ss^\prime} = \frac{q_sq_{s^\prime}}{4\pi\epsilon_0 a k_{\textrm{B}}T},
    \label{eqn:gamma}
\end{equation}
where $q_s$ is the charge of species $s$, $a=(3/4\pi n)^{1/3}$ is the average interatomic spacing
or the Wigner-Seitz radius, and $T$ is the system temperature. This measures the 
ratio of the average potential energy of the system against the average thermal 
kinetic energy. Second, the degeneracy parameter is defined as
\begin{equation}
    \Theta = \frac{k_{\textrm{B}}T}{E_{\mathrm{F}}},
    \label{eqn:theta}
\end{equation}
where $E_{\mathrm{F}} = \hbar^2(3\pi^2 n_e)^{2/3}/(2m_e)$ is the electronic Fermi energy and 
$n_e$ is electron density. %$E_{\mathrm{F}} = \frac{\hbar^2}{2m_e}\ab(3\pi^2 n_e)^{2/3}$
%where $n_e$ is the density of electrons in the system. 
This parameter elucidates 
when the electron-electron interactions become dominantly quantum mechanical. Finally,
\begin{equation}
    r_s = \frac{a}{a_0} =1.8 \frac{q^2}{4\pi\epsilon_0 a E_{\mathrm{F}}}
    \label{eqn:rs}
\end{equation}
is defined as the Coulomb coupling parameter in the degenerate regime. 
Here, $a_0 = 4\pi \epsilon_o\hbar^2/me^2$ is the Bohr radius. 
Once $\Theta<1$, the relevant kinetic
energy of the system is no longer the thermal average and is replaced by the 
Fermi energy.

%Typically, classical plasma physics approaches expand about $\Theta \gg 1 $ and 
%$\Gamma \ll 1$, whereas solid state theories  expand about $\Theta \ll 1$
%and $\Gamma \gg 1$. Warm dense matter sits directly around the point where $\Gamma = 1$ and 
%$\Theta =1$, which means no clear expansion parameter exits to build a theory.

Warm dense matter occurs when all these parameters ($\Gamma$, $\Theta$,  and $r_s$) are of order unity. By 
definition then, it is impractical to develop a theory for warm 
dense matter which expands about any of these parameters. Classical plasma physics
theories tend to follow two formalisms: binary scattering \cite{Spitzer, LeePOF1984, LiPRL1993, BrownPhysRep2005}
and linear response \cite{Nicholson}. These theories fall in the 
classical non-degenerate regime of figure~\ref{fig:phase_space}, where $\Gamma \ll 1$ and 
$\Theta \gg 1$, and are not suited for warm dense matter. An alternative is to use classical molecular dynamics \cite{ZwicknagelLPB1995, GrabowskiPRL2013, BernstienPOP2019},
where Newton's equations of motion are solved on a set of particles. In principle,
this should be valid for all classical areas of figure~\ref{fig:phase_space} ($\Theta \gg 1$),
but is computationally expensive. On the other end of the spectrum, condensed matter based theories 
are based on methods treating the strongly coupled degenerate regime of figure~\ref{fig:phase_space}
where $\Gamma \gg 1$ and $\Theta \ll 1$. The leading technique in condensed matter
is time-dependent density functional theory (TDDFT) \cite{SchleifePRB2015, MagyarCPP2016, DingPRL2018, WhitePRB2018, Whitejop2022, Kononovnpj2023, Kononovpop2024}, which attempts to solve the quantum N-body problem for a set of atoms and a projectile. 
In principle, TDDFT can simulate all conditions, 
but at low densities and high temperatures the number of quantum states necessary is
so large that the problem becomes intractable. As such, the need for intermediate theories that attempt 
to bridge the gap between TDDFT and the traditional plasma 
physics regime is needed. Approaches of this type tend to extend plasma theories either 
through effective potentials in binary scattering theories~\cite{Zwicknagellpb2009, BaalrudPRL2013, BaalrudPOP2019, BernstienPOP2019}
or through the use of local field corrections and structure factors in linear
response theory~\cite{ZwicknagelPhysRep1999, Zwicknagellpb2009, BernstienPOP2019}.
To include the quantum effect of diffraction in particle scattering, a quantum cross section can be considered in place of
a classical one~\cite{GerickePRE1999, GerickePRE2002}.
Additionally, the stopping power resulting from an expression similar to these 
has been integrated over the course of a particle trajectory to calculate its net 
energy exchange with the plasma \cite{GerickePRE2003, CayzacPRE2015}.
Finally, the free electron gas can be considered in place of 
the ideal gas, and similar to the classical non-degenerate regime, both binary 
scattering~\cite{DaligaultPOP2016, DaligaultPOP2018, ShafferPRE2020, RightleyPRE2021} and linear 
response~\cite{Lindhard1964, MerminPRB1970, MelhornJAP1981, WangPOP1998,  MoldabekovPRE2020, HentschelPOP2023}
formalisms for the free electron gas have been developed.

% include paragraph describing what allows BUU theory to extend to these areas
The present theory is an extension of a Boltzmann kinetic equation, similar to 
traditional plasma kinetic theory, but which utilizes the mean force construct~\cite{BaalrudPRL2013,BaalrudPOP2019} 
to extend it to stronger coupling, and the BUU equation to account for degeneracy 
in the particle statistics~\cite{UehlingPRL1933,UehlingPRL1934,RightleyPRE2021}. 
%binary scattering theory which extends to larger coupling 
%and higher degeneracy through the explicit accounting of particle statistics and 
%the potential of mean force.
Electrons are treated as a Fermi gas, so they obey Fermi-Dirac statistics, and 
are treated with full quantum scattering. Additionally, the cross sections are computed based on the  potential 
of mean force. This is the aspect that extends the model to strong coupling~\cite{BaalrudPRL2013,BaalrudPOP2019}. Mean force kinetic 
theory \cite{BaalrudPOP2019} found that the potential of mean force is the correct
interaction potential to preserve the exact equilibrium of a classical system. It is 
interaction force for a binary interaction that accounts for the presence of 
the other $N-2$ particles in the system, treating them in the equilibrium limit. 
In this way, the static influence of many-body physics can be included in binary 
interactions. With the potential of mean force, Boltzmann's molecular 
chaos approximation \cite{Boltzmann} can be relaxed, allowing the regime of plasma kinetic 
theory to be extended from $\Gamma \lesssim 0.1$ 
to $\Gamma \lesssim 30$~\cite{BaalrudPRL2013,BaalrudPOP2019}. 

The mean force method is then extended to quantum systems through the quantum potential
of mean force defined in Ref.~\cite{StarrettHEDP2017}. This method couples the typical
Ornstein-Zernike equation to an average atom calculation \cite{StarrettPRE2013} treating
the electrons quantum mechanically. Along with quantum scattering and the inclusion 
of Fermi-Dirac statistics, the quantum potential of mean force extends the 
range of applicability of the theory from $\Theta \gtrsim 10$ to 
$\Theta \gtrsim 0.1$~\cite{DaligaultPRL2016}. 
The inclusion of these aspects of quantum physics allows the method
to bridge the gap between traditional plasma and solid state regimes. Specifically, 
in the limit $\Theta \gg 1$, it is shown that classical plasma physics results are recovered.
In the opposite limit $\Theta \ll 1$, results of the zero temperature Fermi gas 
are recovered. 

These ideas have been explored previously with the quantum Landau-Fokker-Plank equation~\cite{ShafferPRE2020},
which takes a small scattering angle approximation. Since this is motivated by a weak coupling 
approximation, it makes the application of theory to higher coupling unclear.
The present theory uses a Boltzmann type operator that includes all strengths 
of scattering. 
The crossover from classical to Fermi liquid behavior was also explored by Daligault~\cite{DaligaultPRL2017} 
through a calculation of the momentum lifetime of an electron, showing the transition 
from the weakly coupled classical plasma limit where the Coulomb logarithm arises, 
to the Fermi degenerate limit where the scaling laws of Fermi liquid theory are returned. 
A similar transition is observed here based on an electron-ion collision rate obtained from the low-speed limit of the stopping power. 

%Typically to calculate stopping power for a warm dense matter system, either an analytic formula for region 1 is used 
%or a density functional theory molecular dynamics (DFT-MD) simulation is run, 
%which is more well suited for region 3. Both of these methods have significant 
%draw backs, the region 1 theories contained uncontrolled approximations and 
%are not accurate at all for warm dense matter, whereas DFTMD is extremely accurate, but at
%the relatively high temperatures of warm dense matter, often become extremely computationally 
%expensive. Often these values are used in large scale hydrodynamics codes, which
%need to value speed in this situation, and often reach for the region 1 theories
%or limited tabulated values from DFTMD.

In ICF modeling, stopping power is an input to 
large and computationally expensive hydrodynamic simulations. Commonly, an analytic formula best 
suited for the classical weakly coupled regime is used \cite{Spitzer, LiPRL1993, BrownPhysRep2005}. This is due to the ease 
of calculation, as these theories have a closed form solution and can be evaluated
nearly instantaneously during the simulation. However, past the classical theory limit shown in figure~\ref{fig:phase_space}, the approximations 
within these expressions begin to fail in an uncontrolled way. This can, for example, lead to unphysical  
predictions such as negative collision frequencies. Alternatively, tabulated values from TDDFT simulations
can be used as input, but these calculations tend to be extremely computationally expensive~\cite{Kononovpop2024} 
at the relatively high temperatures characteristic of warm dense matter. This causes these simulations to
take on the order of days to complete, which limits their practicality for exploring the broad range of material
conditions relevant to ICF. The approach presented here, while an 
extension of the classical weakly coupled theories, contains controlled approximations 
valid in the warm dense matter regime. The cost of this is a larger computation time when 
compared to the analytic theories, but still far less than TDDFT. Thus, tables of stopping powers for 
a large array of densities, temperatures, and projectile energies can be computed using the present method
in a fraction of the time it would take for a method based on TDDFT.

The paper is outlined as follows. In section~\ref{sec:theory}, the kinetic 
theory is outlined and an expression for the stopping power is derived. In 
section~\ref{sec:eval}, the expression is evaluated and compared to other leading 
theories. Next, section~\ref{sec:low_speed} examines the asymptotic limits of the stopping power 
and the transition between classical and degenerate behavior is discussed.
Finally, in section~\ref{sec:Barkas}, the Barkas effect~\cite{BarkasPRL1963} in
degenerate plasmas is commented on.

\section{Theory \label{sec:theory}}

\subsection{Boltzmann-Uehling-Uhlenbeck Equation}
The BUU equation is a semi-classical generalization of the classical Boltzmann equation. Its 
collision term is given by \cite{UehlingPRL1933}
\begin{align}
    \mathcal{C}_{ss^\prime}= \int \odif[3]{p_{s^\prime}} & \odif{\Omega} \odv{\sigma}{\Omega} u \Big[ \hat{f}_s \hat{f}_{s^\prime} (1+\delta_s\theta_s f_s)(1+\delta_{s^\prime}\theta_{s^\prime} f_{s^\prime}) \nonumber \\
                                                                                     & -f_s f_{s^\prime} (1+\delta_s\theta_s \hat{f}_s)(1+\delta_{s^\prime}\theta_{s^\prime} \hat{f}_{s^\prime})\Big],
                                                                                     \label{eqn:Boltzmann-Uehling-Uhlenbeck}
\end{align}
where $u = \vab{\bm{p}_s/m_s - \bm{p}_{s^\prime}/m_{s^\prime}}$ is
the relative speed of the collision, $\indv[\sigma]{\Omega}$ 
is the differential scattering cross section, $\theta_s = (2\pi\hbar)^3/g_s$
is the phase space volume occupied by each electron, $g_s$ is the 
spin multiplicity of the species $s$, and 
$\delta_s = -1, 0, 1$ for Fermi-Dirac, Maxwell-Boltzmann, and Bose-Einstein statistics
respectively for species $s$. As in the classical Boltzmann equation, hatted 
quantities denote they are taken post collision, and unhatted quantities are 
taken before a collision. In this work, the differential scattering 
cross section, $\indv[\sigma]{\Omega}$, is determined using the potential of 
mean force which will be defined in section~\ref{sec:PMF}.

This collision operator describes the evolution of the Wigner function, which is 
a generalization of the classical velocity distribution function and a 
representation of the density matrix of quantum statistical mechanics~\cite{Bonitz}. Overall, the BUU 
collision operator can be derived from the Heisenberg equation of motion for a
quantum system in an analogous way to how the classical Boltzmann equation is derived from
the Liouville equation~\cite{BoerckerAOP1979}. This includes an approximation similar
to molecular chaos, which amounts to weak coupling. This means the BUU operator inherits a binary 
scattering approximation in which only two-body interactions are explicitly solved for.  The equilibrium 
state of the BUU collision operator can be shown to be the Fermi-Dirac or Bose-Einstein distribution.
Since only electrons will be treated quantum mechanically here, and are fermions, the Fermi-Dirac 
distribution will be used. This is defined as 
\begin{equation}
    f_s(\bm{p}) = \frac{1}{\theta_s} \frac{1}{\exp(\frac{\beta\ab(\bm{p}-m_s\bm{V}_s)^2}{2m_s} - \beta\mu_s) + 1},
    \label{eqn:FD}
\end{equation}
where $\bm{V}_s$ is the flow shift of species $s$ with respect to an inertial 
reference frame, $\beta=1/(k_{\textrm{B}}T)$ is the inverse temperature, and $\mu_s$ is the chemical
potential of species $s$. The chemical potential is uniquely determined by the 
normalization condition $\int d^3p\, f_s(\bm{p}) = n_s$, which implies
\begin{equation}
    \mathcal{Q}_{1/2}(\beta\mu) = \frac{4}{3\sqrt{\pi}} \Theta^{-3/2},
    \label{eqn:betamu_norm}
\end{equation}
where $\mathcal{Q}_{1/2}(z)$ is a Fermi-Dirac integral of order 1/2. The Fermi-Dirac
integral of arbitrary order is defined as 
\begin{equation}
    \mathcal{Q}_{j}(x) = \frac{1}{\Gamma(j + 1)}\int_0^\infty \frac{y^{j}}{e^{y-x} + 1} \odif{y},
    \label{eqn:FDInt}
\end{equation}
where $\Gamma(x)$ is the gamma function.

The most glaring difference between Eq.~(\ref{eqn:Boltzmann-Uehling-Uhlenbeck}) and the classical Boltzmann
equation are factors in the form of $(1+\delta\theta f)$. In the case of electrons, 
or any fermion, these are Pauli-blocking terms. When a fermion undergoes a binary collision,
the final state of the fermion after the collision 
cannot be occupied. Thus, the term $(1+\delta\theta \hat{f})$ tracks the available states a 
particle can occupy once the collision has completed. Classically, $\delta =0$
and this effect disappears, as all states are equally probable.

The final difference to note between Eq.~(\ref{eqn:Boltzmann-Uehling-Uhlenbeck}) and the classical Boltzmann 
equation is the evaluation of the differential scattering cross section. Since
scattering in warm dense matter is not always classical, a quantum description 
of scattering is used. The scheme used is based on the procedure outlined in Ref.~\cite{LinPPFC2023}.
Since the model is based on a binary interaction assumption, Pauli-blocking does not influence the calculation of the cross section. 
It is accounted for by the weighting factors ($1+\delta \theta f$) described above. 
Details of this calculation are given in Appendix~\ref{app:scattering}.

\subsection{Potential of Mean Force \label{sec:PMF}}

The major result of classical mean force kinetic theory~\cite{BaalrudPRL2013, BaalrudPOP2019}
is that collective interactions in a plasma can be modeled using an effective 
potential in the kinetic equation. The correct potential for a classical system has been rigorously shown 
to be the potential of mean force~\cite{BaalrudPOP2019}. Formally, the potential of mean 
force is the force between 2 particles held at fixed positions obtained by averaging over the 
remaining $N-2$ particles in thermodynamic equilibrium. It is often calculated using the Ornstein-Zernike 
equation along with a corresponding closure equation~\cite{Hansen}. For Coulombic 
systems, the hyper-netted chain approximation is often an accurate closure,
which results from the assumption that the difference between direct and indirect 
correlations are sufficiently small~\cite{Hansen}. In quantum statistical mechanics,
an analogous averaging procedure can be done~\cite{Bonitz}, but the factoring of 
positions and momenta necessary to fix the positions of 
two particles cannot be done. Thus, a quantum analog to Ref.~\cite{BaalrudPOP2019} has not been attempted.  

\begin{figure}
    \includegraphics[width=0.48\textwidth]{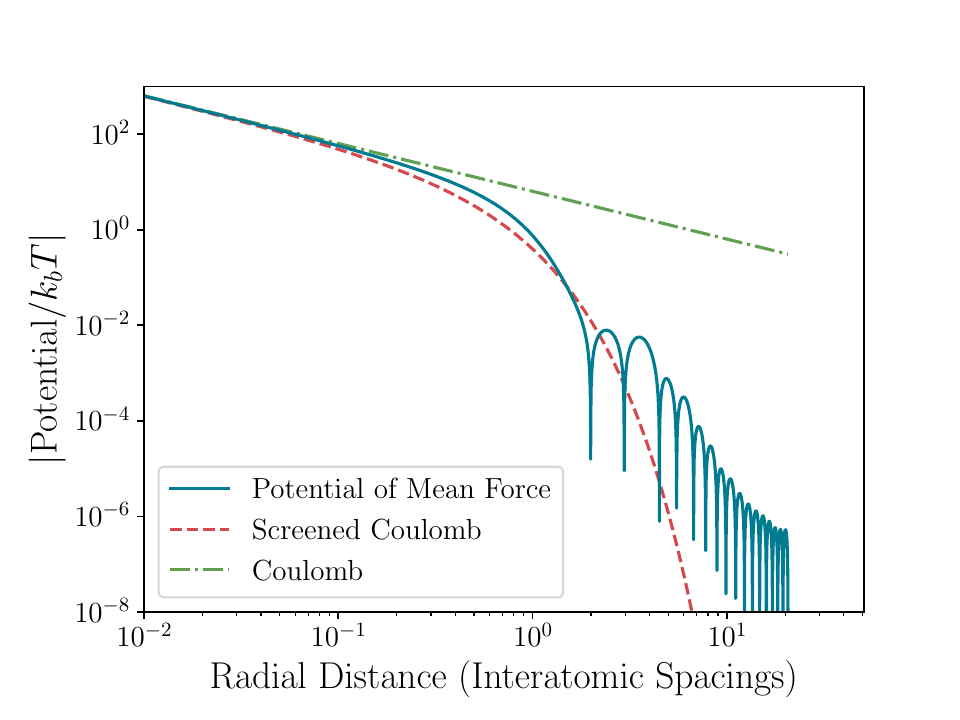}
    \caption{\label{fig:pmf} Scattering potentials for hydrogen at 5eV and 
        1.67~g/cc. This translates to $\Gamma=4.65$ and $\Theta=0.14$, putting this example firmly in the 
        warm dense matter regime. The blue line is the potential of mean force calculated from 
        the average-atom two-component plasma model \cite{StarrettPRE2013, StarrettHEDP2017}.
        The red dashed line is a screened Coulomb potential with a screening length 
        determined by a combination of the Debye length and the Thomas-Fermi
        length~\cite{StantonPRE2016}. The green dashed line shows the Coulomb potential. Notice how the potential
        of mean force contains screening absent from the Coulomb potential,
    and the correlation effects absent from the other potentials.}
\end{figure}

In warm dense matter, ions can still be treated classically \cite{DaligaultPRL2016}.
Due to this, the potential of mean force can be defined for 
ion-ion interactions including screening from degenerate electrons using the average-atom two 
component plasma model~\cite{StarrettPRE2013}. The 
idea is that the electronic structure of the ions needs to be described using an 
average atom (or atom in jellium) model to properly classify electrons as bound
or free. Once this is done, the ions can be coupled together using the Ornstein-Zernike
equation and a quantum hypernetted chain closure, which includes the additional screening degenerate 
electrons provide between the ions. This was further extended to model free
electrons interacting with these average atoms~\cite{StarrettHEDP2017}.  
The definition used here is (given in atomic units $\hbar = m_e = a_0 = e= 4\pi\epsilon_0 = 1$) from Ref.~\cite{StarrettHEDP2017}  
\begin{align}
\label{eqn:V_mf}
    V^{\textrm{MF}}_{ie}(r) = &-\frac{Z}{r} + \int \odif[3]{r^\prime} \frac{n_e^{\textrm{ion}}(r^\prime)}{\vab{\mathbf{r} - \mathbf{r^\prime}}} + V^{\textrm{XC}}\bab{n_e^{\textrm{ion}}(r)}  \nonumber \\
                     & + n_i^0 \int \odif[3]{r^\prime} \frac{C_{ii}\left( \vab{\mathbf{r}-\mathbf{r^\prime}} \right)}{-\beta}h_{ii}\left(\mathbf{r^\prime}\right) \nonumber \\
                     & + \overline{n}_e^0 \int \odif[3]{r^\prime} \frac{C_{ie}\left( \vab{\mathbf{r}-\mathbf{r^\prime}} \right)}{-\beta}h_{ie}\left(\mathbf{r^\prime}\right),
\end{align}
where $Z$ is the atomic number of the ionic species, $n_e^{\textrm{ion}}$ is the density of 
bound atoms around an average atom, $n_i^0$ is the number density of ions in the
plasma, $\overline{n}_e^0$ is the average density of free electrons in the 
plasma, $C_{ii}$ and $C_{ie}$ are the direct correlation functions for ion-ion and 
ion-electron interactions, $h_{ii}$ and $h_{ie}$ are ion-ion and ion-electron 
pair correlation functions, and $V^{\textrm{XC}}$ is the exchange correlation 
potential. 
The choice of $V^{\textrm{XC}}$ is determined by the system in question. Here, 
the Dirac exchange potential~\cite{Dirac1930} (local density approximation) is used.

All quantities in Eq.~(\ref{eqn:V_mf}) are determined self-consistently through the average atom two component plasma model \cite{StarrettPRE2013}.
It is important to note that within the average atom two component plasma model, the average ionization $\overline{Z}$ is 
determined by the spatial integral over the free electron density, defined 
as the electrons with positive energy. Within the model, the point of zero energy is self-consistently determined. Thus, any continuum lowering due to strong correlations is included in the free electron density. In this manner, the ionization, and thus 
electron density, can be a non-integer value, which is not physical for a single isolated
ion. However since the model solves for the average atom of a system, this is a good 
approximation of the mean ionization~\cite{StarrettPRE2024}. 

Overall, the potential of mean force includes multi-particle correlations which are not present 
in most scattering potentials. The correlations manifest in oscillations and 
correspond to average particle positions in the plasma. This can be seen in Fig.~\ref{fig:pmf}, which 
shows various choices for scattering potentials in hydrogen at 1.67g/cc and 5eV.
These density and temperature conditions translate to Coulomb coupling coupling
strength of $\Gamma = 4.65$ and a degeneracy parameter of $\Theta=0.14$ corresponding to the
warm dense matter regime. 
%The solid blue line is the ion-electron potential of mean force calculated using 
%Ref.~\cite{StarrettHEDP2017}. The red dashed line is a screened coulomb 
%potential with a screening parameter common to the warm dense matter regime \cite{StantonPRE2016},
%and the green dashed-dotted line is the classic coulomb potential. Note how the 
%coulomb potential contains neither the screening physics nor the correlation 
%physics the potential of mean force does. Additionally, the screened coulomb potential seems to 
%qualitatively capture the screening physics, but does not capture any of the 
%correlation physics (seen by the oscillations) present in the potential of mean force
This is why
mean force kinetic theory, and hence the present theory, can extend to higher coupling. It is due to the 
additional screening and correlations that the potential of mean force captures. 

\subsection{Stopping Power}
With these ideas, an expression for stopping power can be developed.
Following Ref.~\cite{RightleyPRE2021}, the stopping power is computed from the friction 
force density between two species in the plasma
\begin{equation}
    \bm{R}_{ss^\prime} = \int \odif[3]{p_s}\bm{p_s}\mathcal{C}_{ss^\prime}.
    \label{eqn:friction}
\end{equation}
As with classical collisional processes, the largest momentum transfer occurs
between particles of similar average speeds. In an 
equilibrium plasma the ratio between the speed of ions and the speed of electrons 
is $v_{\textrm{av},e}/v_{\textrm{av},i} \approx \sqrt{m_i/m_e} \sim 43$, at the least, due to their relative mass difference.
In ICF, fusion alpha particles are born at 3.5 MeV~\cite{BettiJOP2016}.
This means that thermal electrons provide the largest momentum transfer,
and hence stopping power, at a thermal energy of around 1 keV. If instead the alpha particles were to stop
on thermal protons, the maximal momentum transfer would require the protons to 
have a thermal energy of 100s keV. As such, the only contribution considered here is of ballistic ions stopping on free electrons. The projectile
ion species $s'$ is approximated as a beam with velocity $\bm{V}_i$. As such, its velocity distribution function is given by $f_i(\bm{p}) = n_i \delta\ab(\bm{p} -m_i \bm{V}_i)$,
where $n_i$ is the density of ballistic ions, and $\delta$ is the Dirac delta 
function. The free electrons follow the Fermi-Dirac distribution given in 
Eq.~(\ref{eqn:FD}), and have a free electron density $n_e$ determined by the average 
ionization, $\overline{Z}$, from the average atom two component plasma model. 

Additionally, two simplifications can be made to the integrals of Eq.~(\ref{eqn:friction}).
The first takes advantage of the mass ratio already used to isolate electronic stopping 
power. When an electron and ion undergo a binary collision, the change in velocity of the ion 
is much smaller than that of the electron. This means it is a good approximation 
to take the limit $m_r = m_e/m_i \ll 1$ in order the simplify the collision physics. 
Since the integrals and distribution functions were defined in
terms of momenta above, which includes a factor of the mass, these are changed to velocities here. 
Along with this, it is helpful to normalize all velocities by the electron 
thermal speed, from this point forward $\tilde{\bm{v}}_s = \bm{v}_s/v_{te}$
and $\tilde{\bm{V}}_s = \bm{V}_s/v_{te}$, where $v_{te} = \sqrt{2k_{\textrm{B}}T/m_e}$. 
The second simplification takes advantage of symmetries present in the differential
scattering cross section. It is known that the scattering cross section should
be symmetric with respect to time reversal and space inversion~\cite{Sakurai}.
This means that the pre and post collision velocities in the collision operator can be exchanged
and the expression will still hold. This is a manifestation of the principle of 
detailed balance for a binary, elastic collision~\cite{Boltzmann, Reif, Cercignani, FK}. 

Using these simplifications, Eq.~(\ref{eqn:friction}) becomes
\begin{align}
    \bm{R}_{ie} = -&\frac{m_en_in_ev_{te}^2}{\mathcal{Q}_{1/2}(\beta\mu_e)\pi^{3/2}}\int \odif[3]{\tilde{u}}\odif{\Omega} \odv{\sigma}{\Omega}\nonumber \frac{\tilde{u}\bm{\Delta \tilde{u}}}{e^{\ab(\tilde{\bm{u}} + \bm{\Delta \tilde{V}})^2-\beta\mu_e}+1} \nonumber\\
                         &\times \ab[\frac{e^{\ab(\tilde{\bm{u}}+\bm{\Delta \tilde{u}}+\bm{\Delta\tilde{V}})^2-\beta\mu_e}}{e^{\ab(\tilde{\bm{u}}+\bm{\Delta \tilde{u}}+\bm{\Delta\tilde{V}})^2-\beta\mu_e}+1}],
                         \label{eqn:pre_ref}
\end{align}
where $\bm{\Delta \tilde{u}} = \ab(\hat{\bm{u}} - \bm{u})/v_{te}$ is 
the change in the relative velocity vector $\bm{u}$, and $\bm{\Delta \tilde{V}} 
= \tilde{\bm{V}}_i - \tilde{\bm{V}}_e$ is the velocity of the ballistic ions in 
the center of mass reference frame of the electrons. The expressions of Ref.~\cite{RightleyPRE2021}
can be obtained by taking the limit $|\bm{\Delta \tilde{V}}| \ll 1$ in 
Eq.~(\ref{eqn:pre_ref}). Stopping power is a drag force, which is obtained from 
the friction force density as $\bm{R}_{ie}/n_i$. Since it acts antiparallel to the ion velocity
vector, the relationship between $\bm{R}_{ie}$ and 
stopping power is given by
\begin{equation}
    -\odv{E}{x} = \frac{\bm{\Delta V}}{\Delta V} \cdot \frac{\bm{R}_{ie}}{n_i},
    \label{eqn:stoppingpower}
\end{equation}
where $\indv[E]{x}$ is the stopping power.

It is clear that Eq.~(\ref{eqn:pre_ref}) must be evaluated numerically. 
However, further simplifications can be made.
First the integral is put into spherical coordinates and a reference frame is chosen
such that the only free parameter, $\bm{\Delta V}$, is oriented along the cartesian
z-axis. Details of the explicit vector
components are given in Appendix \ref{app:ref}. Once the transformation into this reference frame is 
substituted into Eq.~(\ref{eqn:pre_ref}), it is then a simple math exercise to 
reduce the expression into the form
\begin{widetext}
    \begin{align}
        \bm{R}_{ie} =&\frac{2m_en_in_ev_{te}^2}{\mathcal{Q}_{1/2}(\beta\mu_e)\pi^{1/2}} \frac{\bm{\Delta\tilde{V}}}{\Delta\tilde{V}} \int \odif{\tilde{u}} \odif{\theta^\prime} \odif{\theta} \odif{\tilde{\phi}}\, \sin\theta^\prime  \sin\theta  \odv{\sigma}{\Omega}
        \tilde{u}^4\frac{(1-\cos\theta)\cos\theta^\prime+\sin\theta^\prime\sin\theta\cos\tilde{\phi}}{e^{\tilde{u}^2 + \Delta\tilde{V}^2 +2u\Delta\tilde{V}\cos\theta^\prime -\beta\mu_e} + 1} \nonumber \\
                     &\times\ab(1 - \frac{1}{e^{\tilde{u}^2 + \Delta\tilde{V}^2 +2u\Delta\tilde{V}g(\theta, \theta^\prime, \tilde{\phi}) -\beta\mu_e} + 1}),
                     \label{eqn:final}
    \end{align}
\end{widetext}
where $\theta^\prime$ is the polar angle for the spherical integration of $\tilde{\bm{u}}$,
$\theta$ is the polar scattering angle, $\tilde{\phi} = \phi-\phi^\prime$, where 
$\phi^\prime$ and $\phi$ are the azimuthal angles for the integration of $\tilde{\bm{u}}$
and the scattering angle respectively. Finally $g(\theta, \theta^\prime, \tilde{\phi}) =
\cos\theta\cos\theta^\prime - \sin\theta\sin\theta^\prime\cos\tilde{\phi}$ is an
angular factor describing the angle between $\bm{\Delta \tilde{u}}$ and $\bm{\Delta \tilde{V}}$.

In comparison to Eq.~(\ref{eqn:pre_ref}), Eq.~(\ref{eqn:final}) is simpler to evaluate 
because it is a 4-dimensional rather than 5-dimensional integral, and only one 
of the integrals is improper. Despite this, the integral does display some challenges since
the distribution functions depend explicitly on all angular integration variables. This means this integral cannot be 
separated any further and the differential 
scattering cross section cannot be reduced into a momentum transfer or energy transfer
cross section. 

It is important to state that
due to the use of the average atom model, only unbound electrons are taken to participate
in the stopping. This is reasonable if the system in question has pressure ionized 
all but a few core electrons, or if the system is not dense at all. This makes various 
forms of hydrogen and lower $Z$ elements good candidates for this theory. 
Additionally, since the atomic structure is not being explicitly solved for,
inelastic processes such as bound-free transitions are not solved for.

\section{Results \label{sec:eval}}

\begin{figure*}[t]
    \includegraphics[width=0.98\textwidth]{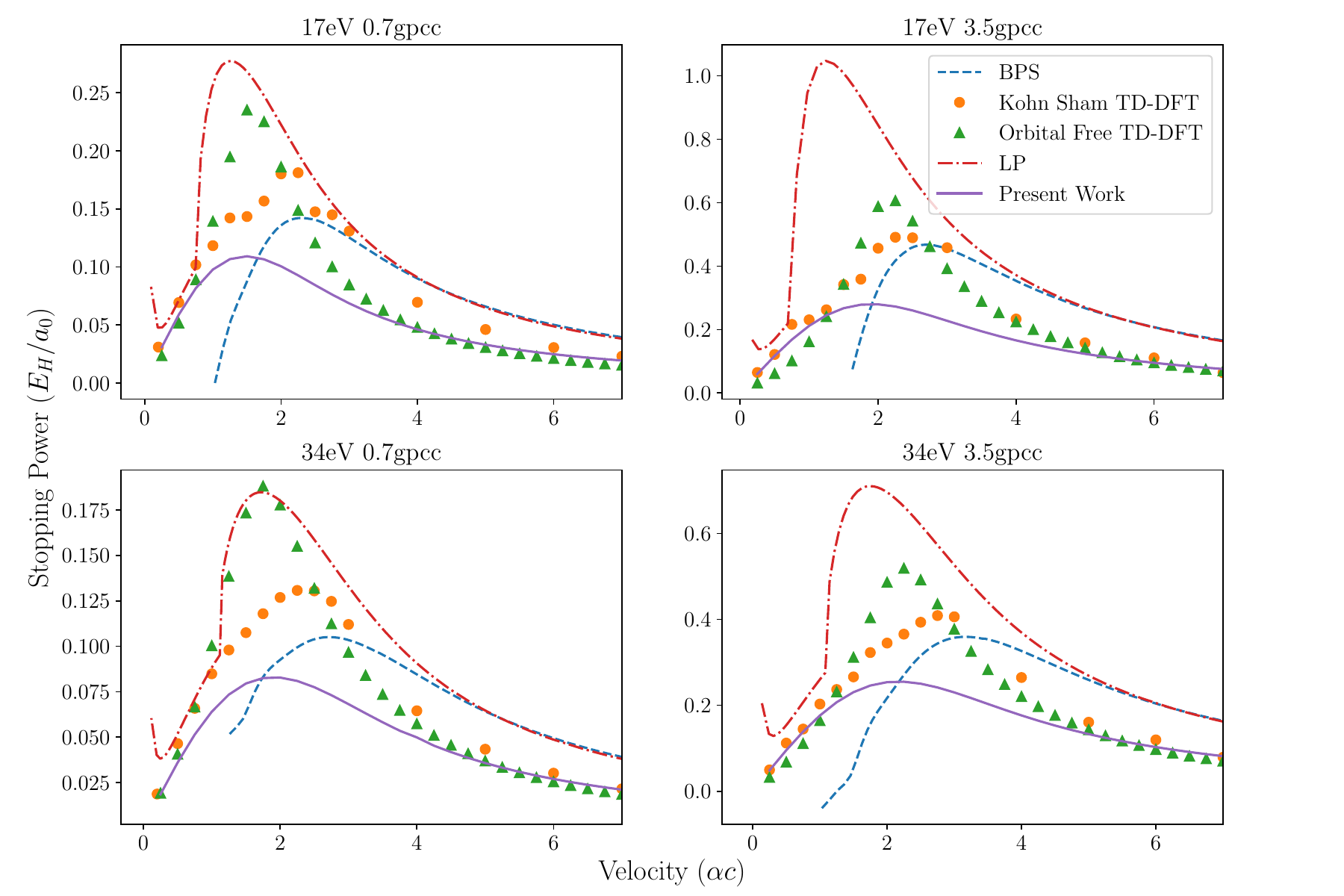}
    \caption{\label{fig:white_comp} Stopping power of a deuteron 
    in warm dense deuterium at various densities and temperatures. The results of the 
    present method are shown as solid purple lines. 
    This is compared to two analytic theories: Brown-Preston-Singleton~\cite{BrownPhysRep2005} (dashed blue lines) and 
   Li-Petrasso~\cite{LiPRL1993} (dashed-dotted red lines). We also compare to orbital-free (green triangles) and
    Kohn-Sham (orange circles) TDDFT calculations~\cite{WhitePRB2018}. Our 
    method matches the density functional theory  calculations in both the low speed and high speed limits.
    Neither analytic theory is consistent with both the high and low velocity limits of 
    the density functional theory, as they are not well 
    suited for the warm dense matter regime. Units are given in atomic units where
    $E_\textrm{H}=27.2$eV is the Hartree energy, $a_0$ is the Bohr radius, and $\alpha=e^2/(4\pi \epsilon_o \hbar c)$ is the 
    fine structure constant.}
\end{figure*}

As a test case, we consider deuterons stopping in warm dense deuterium. This 
was chosen because it is relevant to ICF 
implosions \cite{HainesPOP2024}, and because there is previous calculation data to compare with. Specifically, the results are compared to
two analytic theories~\cite{LiPRL1993, BrownPhysRep2005} from the classical 
weakly coupled region of Fig.~\ref{fig:phase_space}, and two 
formalisms of TDDFT molecular dynamics simulations~\cite{WhitePRB2018}.
Time dependent density functional theory is a quantum mechanical technique to 
simulate many body quantum systems. In the results compared against, electronic wave functions
were time evolved in two different ways. The Kohn-Sham formalism solves for wave functions 
using a time-dependent extension to the Kohn-Sham equation~\cite{DFT1, DFT2} along 
with the Mermin extension for finite temperature~\cite{MerminPhysRev1965}. As such,
TDDFT carries similar assumptions and approximations as finite temperature time-independent DFT, for example the exchange-correlation potential or the initial distribution of 
occupied states. Since the comparison is done with a simple element, deuterium, 
it is expected that Kohn-Sham TDDFT should be a good approximation to the system,
as it has been used in many other warm dense matter systems~\cite{BaczewskiPRL2016, Kononovnpj2023, Kononovpop2024}.
This makes it a good benchmark for the present theory. One additional drawback of TDDFT 
is that it is extremely computationally expensive, often requiring extreme computational
resources \cite{Kononovnpj2023, Kononovpop2024}. 
%Time dependent density functional theory is widely reguarded as the gold standard 
%for large quantum systems. Electronic wave functions can be time evolved for hundreds
%of atoms in one simulation, making it one of the most complete physics descriptions 
%for the warm dense matter regime \cite{BaczewskiPRL2016}. In the Kohn-Sham formalism,
%wave functions are solved for using the Kohn-Sham equation \cite{DFT1, DFT2}
%and the Mermin extension for finite temperature \cite{MerminPhysRev1965}. Since 
%the comparison is done with a simple element, deuterium, it is expected that 
%the Kohn-Sham TDDFT should model most, if not 
%all of the physics. This makes it a good benchmark for the present theory. The cost 
%of including so much physics is that this method becomes extremely computationally 
%expensive. 
It has been shown to scale as the cube of temperature~\cite{WhitePRB2018},
making convergence in the temperature regimes of warm dense matter difficult. 

Figure~\ref{fig:white_comp} shows that for the cases chosen, the BUU theory matches the 
Kohn-Sham TDDFT fairly well, with the largest deviations occurring near the Bragg peak. They agree especially well at low projectile speeds where the 
stopping power can be related to linear transport process such as electrical 
conductivity. Agreement between these two methods in this regime has been shown before \cite{RightleyPRE2021}.
The surprising agreement comes at higher velocities. At high projectile speeds, it is standard
to use velocity dependent potentials, known as dynamic screening, to capture the linear 
response of the plasma to the projectile~\cite{MelhornJAP1981, WangPOP1998, ZwicknagelPhysRep1999}. Time dependent 
density functional theory can self consistently solve for this, but the potential 
of mean force is a static, velocity-independent potential which means that dynamic aspects of the linear
response are not included. The agreement may be due to the fact that in the present 
theory, collisions with electrons are treated fully quantum mechanically. This 
includes the full treatment of quantum scattering in the partial wave expansion, 
and the inclusion of Pauli blocking in the collision operator to properly weight 
collisions.

At projectile speeds around the Bragg peak, the model shows the largest 
disagreement with the Kohn-Sham TDDFT. 
This is also where the different TDDFT methods have the largest scatter in the data, and where they disagree with one another most. 
In Ref.~\cite{WhitePRB2018}, the scatter in the data was largely attributed to finite size effects in the simulation. 
It was also mentioned that the TDDFT data includes effects of bound electrons in 
the projectile particle and that better agreement with dielectric-function based 
models was achieved if the projectile is fully stripped. 
A similar effect may be of some influence here, as any inelastic physics of bound-free 
transitions, which result from dynamic many body effects, can be modeled by TDDFT, but are absent from the present theory. 
% This
% makes a large impact around the Bragg peak when the projectile speed is similar 
% to the average electron speed, and the largest momentum transfer can occur. 

A comparison is also made to orbital free TDDFT.
In this formalism, electrons are treated semi-classically with a finite temperature 
Thomas-Fermi type equation \cite{FeynmanPhysRev1949, WhitePRB2018}. This is not 
expected to capture bound electron physics, but is a good approximation at higher 
temperatures where warm dense matter exists. Its scaling in temperature is not 
as severe as the Kohn-Sham variety, but it still relies on molecular dynamics 
which is computationally expensive on its own \cite{WhitePRB2018}. As such, the orbital free 
density functional theory is consistent with the Kohn-Sham variety and the present 
method away from the Bragg peak, but disagrees at the Bragg peak. 
%This is again likely because
%the treatment of bound electrons is not fully correct in the orbital-free method. 

Finally, two analytic models which are commonly used in ICF applications are compared to. These are the Li-Petrasso \cite{LiPRL1993}
and Brown-Preston-Singleton \cite{BrownPhysRep2005} theories. They are both 
formulated using a Fokker-Plank type equation along with the Coulomb potential, which is not valid at strong 
coupling and hence warm dense matter. In both, a correction is included which 
is attributed to plasma screening. This is a step towards including dynamic 
screening, which could explain the difference in the predicted tail of the these
two theories when compared to either of the TDDFT methods or the BUU method. The more likely cause 
of the disagreement is the Coulomb coupling. All of the cases considered in
Fig.~\ref{fig:white_comp} have a Coulomb coupling parameter $\Gamma > 0.4$. These 
plasmas are too strongly coupled for these analytic theories which explicitly 
expand about $\Gamma \ll 1$ \cite{LiPRL1993, BrownPhysRep2005}. Additionally, 
Li-Petrasso and Brown-Preston-Singleton do not contain any notion of Pauli blocking. 
%which could make a difference at large speeds. 
Pauli blocking limits the energies
electrons can leave a collision with, which means that collisions which these models predict energy transfer from the projectile to the plasma
could actually be blocked. This would lead to an overall reduction in the stopping 
power, which is observed.

\begin{figure}
    \includegraphics[width=0.48\textwidth]{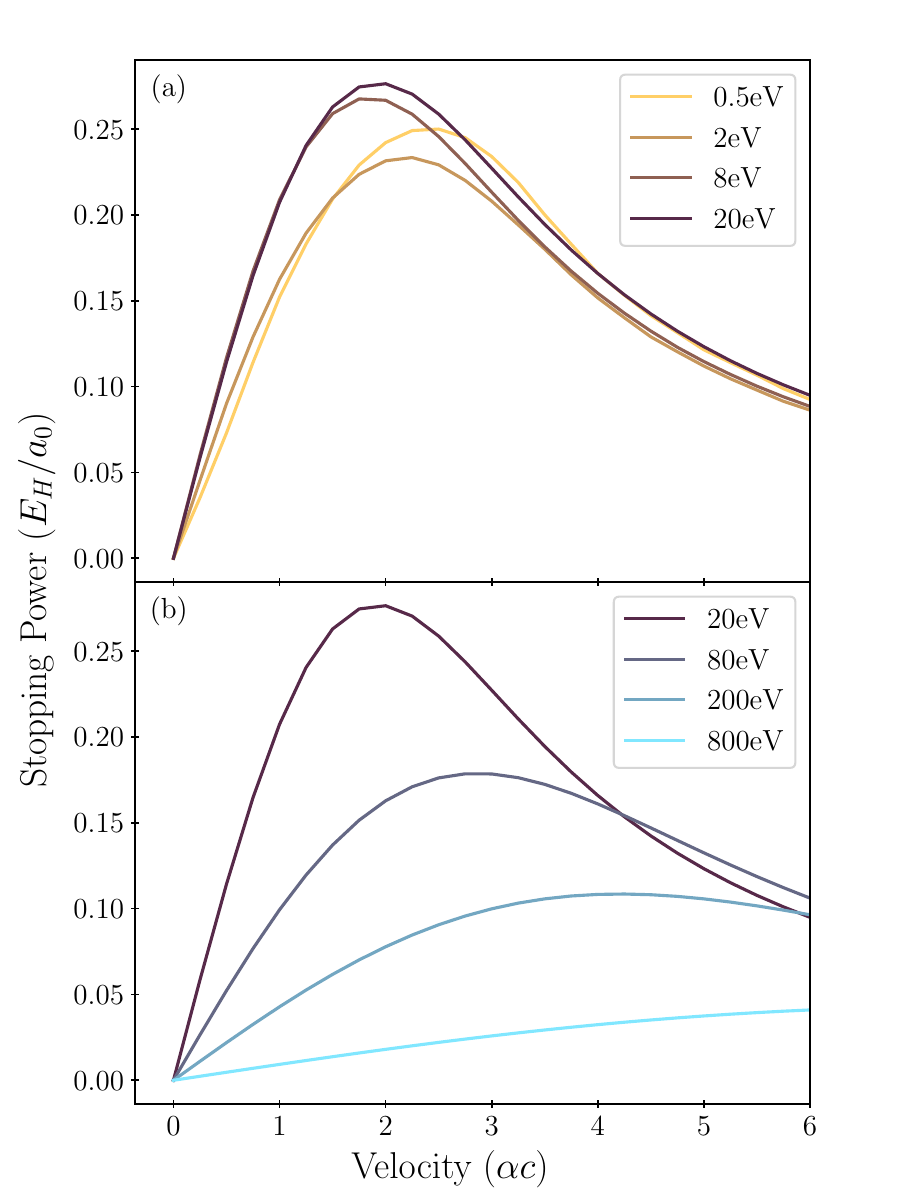}
    \caption{\label{fig:peaks} Predictions of proton stopping power in 
    hydrogen at 1.67 g/cc and various temperatures. Panel (a) shows the 
    results from 0.5eV to 20eV, which are characterized by a degenerate regime, $\Theta <1$.
    Panel (b) shows the results for 20eV to 800eV, which are characterized
    by a classical regime, $\Theta>1$. Notice that the stopping power curves for the degenerate 
    cases in (a) do not change very much over the span of temperatures. This 
    because the ballistic proton is stopping on electrons at the Fermi energy,
    which is temperature independent. Whereas in (b), the curve changes substantially
    with temperature as the proton is stopping on electrons at the thermal velocity,
    which increases with temperature. Units are given in atomic units where
    $E_H=27.2$eV is the Hartree energy, $a_0$ is the Bohr radius, and $\alpha$ is the 
    fine structure constant.} 
\end{figure}

The main benefit of the BUU approach is that it can capture the relevant 
strong coupling and quantum effects relevant to warm dense matter at a fraction of the computational cost. Especially at lower speeds, 
away from velocities which require the need of dynamic screening or inclusion of 
bound-free electron transitions. With this in mind, it is expected this approach should work 
best when there are few bound electrons in the plasma or when the 
bound electrons are highly localized around their nucleus.
%The main benefit to the present approach with the Boltzmann-Uehling-Uhlenbeck equation, is that this method
%can be a good approximation to DFT at a fraction of the computational cost. Our 
%approach will often take on the order of minutes to calculate compared to
%the days or weeks DFTMD takes. Of course this timing cannot compete with the 
%computational cost of analytic theories, but this comes at a great increase in
%accuracy. 

\begin{figure*}
    \includegraphics[width=0.96\textwidth]{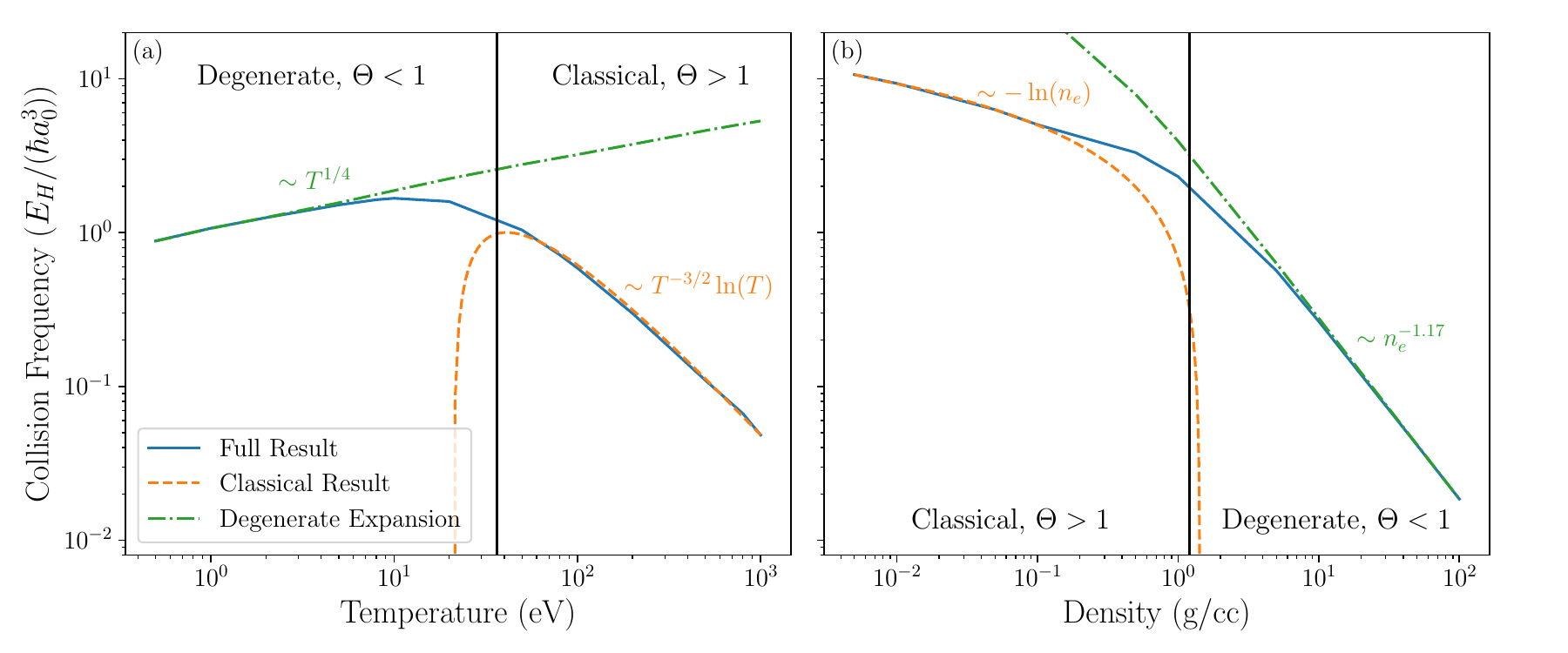}
    \caption{\label{fig:slopes} Electron ion collision frequency 
    of a hydrogen plasma computed using Eq.~(\ref{eqn:coll}). The two plots show the (a) temperature and (b)
    density dependence in both the degenerate and classical limits. Notice 
    that the behavior of the two limits is qualitatively very different, 
    suggesting that the characteristic energies involved in collisions has 
    changed from the thermal speed of the plasma to the Fermi energy.
    The black vertical line on both plots denotes the point $\Theta =1$,
    and represents the location of the transition, as well as where warm dense matter 
    conditions are present. Units are given in atomic units where
    $E_H=27.2$eV is the Hartree energy, $a_0$ is the Bohr radius, and $\alpha$ is the 
    fine structure constant. Note that the collision frequency calculated is actually divided by 
    the ion density, $n_i$, as a density of ions for a ballistic projectile has no 
    physical meaning.}
\end{figure*}

%As a standard for comparing computational methods, some of the cases
%from the Second Charged Particle Transport Coefficient Code Comparison Workshop~\cite{StanekPOP2024} 
%were revisited. In particular, 
Additionally, predictions were made for the stopping of protons 
in hydrogen at 1.67 g/cc at temperatures ranging from 0.5 eV to 1 keV; see Figure~\ref{fig:peaks}. A predicted 
consequence of the Pauli blocking factors in the BUU 
equation is that the stopping power curve is insensitive to temperature when $\Theta \ll 1$. This is 
because collisions are limited to an energy range of width $k_{\textrm{B}}T$ around
the Fermi energy. When the temperature becomes small,  
this energy range collapses so that collisions can only occur at the 
Fermi energy. The Fermi energy is temperature independent, so stopping 
power should not have explicit
temperature dependence in this limit. This expectation is born out in the data shown in Fig.~\ref{fig:peaks}(a), which 
shows the stopping powers for a range of temperatures in the degenerate regime $(\Theta < 1)$.
In contrast, the stopping power curve in the classical regime $(\Theta > 1)$ depends 
strongly on temperature. Here there is no blocking of states, and electrons can 
scatter across the full thermal distribution of energies. A consequence is that 
the Bragg peak will follow the average speed of thermal distribution, the thermal
speed. This can be seen in Fig.~\ref{fig:peaks}(b) which shows stopping powers 
for a range of temperatures in the classical regime.

This behavior signifies there is a transition across the warm dense matter regime
from classical plasma physics to degenerate plasma physics~\cite{DaligaultPRL2017}. This transition is the 
illustration of why $\Theta$ is the proper parameter to characterize the degeneracy of
the system. It signifies when the system transitions from being determined by the 
thermal energy $k_\textrm{B}T$ to the Fermi energy \cite{LampePhysRev1968}. This phenomenon
has even been observed in experiments on the National Ignition Facility~\cite{HayesNature2020}
and the Omega laser~\cite{ZaghooPRL2019}.

\section{Asymptotic Behavior of Stopping Power in Warm Dense Matter \label{sec:low_speed}}

Along with the stopping power, the electron-ion collision frequency is expected 
to display different scalings with respect to density and temperature in the classical and degenerate regimes. Again,
this is because collisions in the classical regime sample the entire temperature range of the Maxwell-Boltzmann distribution,
whereas in the degenerate regime Pauli blocking limits them to around the Fermi energy. This 
fundamentally changes how energy and momentum are transported in degenerate
as opposed to classical plasmas~\cite{DaligaultPRL2017}. 

%To further investigate the transition between a plasma state and a solid-like 
%state, it is illuminating to look at the asymptotic behavior of
%the Bragg curve as a function of the degeneracy parameter, $\Theta$. The most 
%likely candidate for investigation is the Bragg peak, but this quantity can be 
%difficult to identify in sparse data like the time dependent density functional theory provides, and further is not 
%very consistent. With these points in mind, it is better to identify the low 
%speed behavior of the stopping power curve, as this quantity is related to 
%the fundamental electron ion collision rate in a plasma and has been verified 
%against different theories previously \cite{RightleyPRE2021, StanekPOP2024}. 

From the stopping power, the electron-ion collision frequency can be found by 
taking the small projectile velocity limit. Explicitly this is done by taking the limit
$\bm{\Delta \tilde{V}} \ll 1$ in Eq.~(\ref{eqn:final}), which yields
\begin{align}
    \bm{R}_{ie} = -&m_en_e\nu_{ei}\bm{\Delta V},
    \label{eqn:small_V}
\end{align}
where
\begin{align}
    \nu_{ei} = &\frac{8}{3}\frac{n_i v_{te}}{\mathcal{Q}_{1/2}\ab(\beta\mu_e)\pi^{1/2}}\int_0^\infty \odif{\tilde{u}}\sigma_p \tilde{u}^4 G\ab(\tilde{u}^2),
    \label{eqn:coll}
\end{align}
is the electron-ion collision frequency,
$\sigma_p$ is the momentum transfer cross section defined as 
\begin{equation}
    \sigma_p = \int \odif{\Omega} \odv{\sigma}{\Omega} \ab(1-\cos\theta),
    \label{eqn:sigma_p}
\end{equation}
$G\ab(\epsilon)$ is related to the availability of states given by 
\begin{equation}
    G(\epsilon) = \frac{\sqrt{\epsilon}e^{\epsilon-\beta\mu_e}}{(e^{\epsilon-\beta\mu_e} +1)^2} = \sqrt{\epsilon} f_\textrm{FD}(1-f_\textrm{FD}),
    \label{eqn:gu}
\end{equation}
where $f_\textrm{FD}$ is the Fermi-Dirac distribution, and $\epsilon = \tilde{u}^2$ is the dimensionless relative particle energy. 
In comparison to Eq.~(\ref{eqn:final}), Eq.~(\ref{eqn:small_V}) is much simpler to evaluate
as all but one integral can be integrated out or absorbed into other factors. 
The momentum transfer cross section has a 
much simpler form than the differential scattering 
cross section due to angular symmetries. This is shown for quantum scattering in Appendix \ref{app:scattering}. 

When changing the temperature (at a constant density of 1.67 g/cc) and changing 
the density (at a constant temperature of 20 eV) while evaluating Eq.~(\ref{eqn:small_V}),
different behavior exists in the degenerate and classical regimes. This can be 
seen in Fig.~\ref{fig:slopes}, but to make this more evident, explicit expansions in large 
and small $\Theta$ can be performed. 

In the classical limit $\Theta \gg 1$, the availability of 
states $G\left(\epsilon\right)$ reduces to a Maxwellian form, and the classical result is recovered~\cite{Spitzer}
\begin{align}
    \nu^\textrm{C}_{ei} = &\frac{8}{3}\frac{n_i v_{te}}{\pi^{1/2}} \int_0^\infty \odif{\tilde{u}}\sigma_p \tilde{u}^5 e^{-\tilde{u}^2}.
    \label{eqn:classical}
\end{align}
The results from this expression, Eq.~(\ref{eqn:classical}), follow the scalings
in temperature and density of $\nu^{C}_{ei} \sim T^{-3/2}\ln\left(T\right)$ and 
$\nu^{C}_{ei} \sim -\ln\left(n_e\right)$ respectively. This is the expected 
picture from the view of classical plasma physics \cite{Spitzer}.

%kWe see that around 36eV the classical method fails and produces a 
%negative slope, something which is not physical. This is due to the classical 
%coulomb logarithm failing, and the point it does, around 36 eV or the black vertical
%line on Fig~\ref{fig:slopes}, is where $\Theta = 1$. Thus we know that anywhere the 
%classical description is valid, the ballistic ion is stopping on thermal electrons.
%We would like to note that it seems like the classical result should be valid 
%until $\Theta \sim 1$, but that is not always true. In this case it is because at this density, as the 
%temperature decreases, the $\Theta=1$ point is reached before the $\Gamma=1$ point.
%We are moving down a vertical line in Fig.~\ref{fig:phase_space} that is 
%to the right of the $r_s = 1$ line. Conversely, if the density were lower, the 
%vertical line we travel along in fig.~\ref{fig:phase_space} would be to the left 
%of the $r_s=1$ line, and we would reach $\Gamma =1$ first as the temperature decreases.
%This means the classical result would fail far before the point $\Theta = 1$. 

\begin{figure}
    \includegraphics[width=0.48\textwidth]{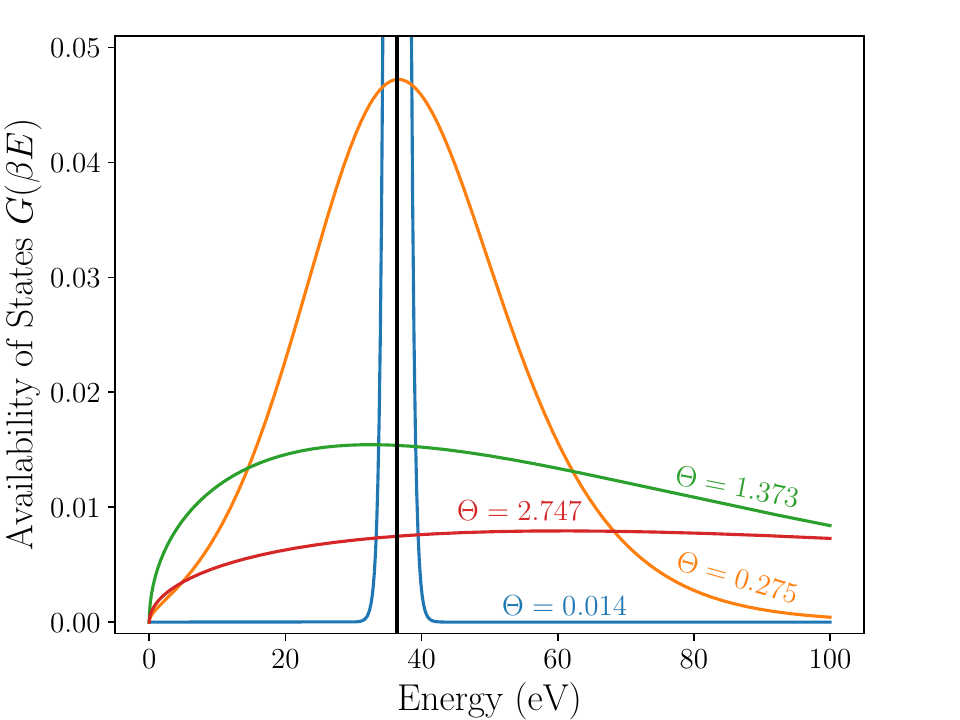}
    \caption{\label{fig:dist}  The availability of states given by Eq.~(\ref{eqn:gu}) is plotted 
        for various values of $\Theta$. Here $\beta$ is the inverse temperature 
        and $E$ is the energy in electron volts. This function can be thought of as related to 
        the occupancy of states in the plasma. Notice that when $\Theta >1$, we seem 
        to recover an exponential trend, similar to a Maxwell-Boltzmann distribution.
        When $\Theta < 1$, the distribution follows a peaked form and begins to become  
        strongly peaked around the Fermi energy denoted by the black vertical line. 
        This suggests that electrons are only able to undergo a collision if they 
    sit around the Fermi energy.}
\end{figure}

In the degenerate limit, $\Theta \ll 1$,  it is not clear how to expand $G\left(\epsilon\right)$ with a simple Taylor expansion. Thus, it is informative
to examine the behavior of $G(\epsilon)$ as a function of $\Theta$. 
% It is also 
% convenient to discuss the prop, $\epsilon = \tilde{u}^2$, which acts as a 
% normalized energy to discuss the properties of $G$. 
In this regime, Pauli blocking 
prevents electrons far from the Fermi energy from participating in collisions. This 
means $G(\epsilon)$ can be approximated as a Dirac
delta function around the Fermi energy to a good approximation as $\Theta \rightarrow 0$. This behavior is illustrated in Fig.~\ref{fig:dist}
for four different values of $\Theta$. Interestingly, this argument is not dissimilar
from those commonly made of continuum electrons in metals \cite{Ashcroft}. 
In fact, this is the motivation for the Sommerfeld expansion \cite{SommerfeldZP1928}.
The expansion involves assuming the distribution function is strongly peaked around a 
point, which Fig.~\ref{fig:dist} shows to be the Fermi energy. Then, a Taylor expansion of the 
remaining part of the integrand around that point is taken. For this case, the 
rest of the integrand includes the cross section, which is not 
known analytically and is therefore difficult to expand.
Alternatively, the lowest order term can be obtained by taking the availability of 
states $G(\epsilon)$ to be a Dirac delta function around the Fermi energy. Doing so leaves the expression
\begin{equation}
    \nu_{ei}^\textrm{Q} = n_i\sqrt{\frac{2E_{\mathrm{F}}}{m_e}}\sigma_p\ab(E_{\mathrm{F}})
    \label{eqn:degen}
\end{equation}
where $\sigma_p\ab(E_{\mathrm{F}})$ is the momentum transfer cross section evaluated
at the Fermi energy. 

A weak temperature scaling is expected in Eq.~(\ref{eqn:degen}). 
Although there is no explicit temperature dependence, the potential of mean force implicitly depends on temperature and is highly dependent on the material of interest. 
If a Coulomb potential were to be used, the expression would be completely independent 
of temperature. This result is fundamentally different than the classical scaling
in both density and temperature, signifying a difference in physics  
due to the protons colliding with thermal electrons in the classical limit 
versus at the Fermi energy in the degenerate limit. Examining these expressions 
closer, notice $\nu_{ei}^\textrm{C} \propto v_{t}$, whereas $\nu_{ei}^\textrm{Q} \propto v_\textrm{F}$, 
where $v_\textrm{F} = \sqrt{2 E_\textrm{F}/m_e}$ is the Fermi speed. These are the average speeds of 
a classical and degenerate gas respectively, further signifying how the statistics
influences the speed at which particles can scatter, and in turn, the characteristic
scaling of the collision frequency. This is seen in Fig.~\ref{fig:slopes}, as the 
scalings in both density and temperature are different in the classical and degenerate 
regimes.

This same behavior can be translated to stopping power, which is why Fig.~\ref{fig:peaks}(b) can be described by classical
plasma physics behaviors, whereas Fig.~\ref{fig:peaks}(a) is dominated by collisions at the 
Fermi energy.
This is interesting because the presented theory is in principle a high temperature 
gas theory, but has qualitative similarities with some solid-state theories
at lower temperatures. The overall accuracy of the calculations is questionable 
outside of the warm dense matter regime, since it is not expected that the necessary correlations are 
captured. Despite this, it is interesting to see the qualitative trends 
which seem to match scaling laws of Fermi liquids.

\section{Comment on the Barkas Effect \label{sec:Barkas}}

In classical plasma physics, collisional transport properties, including stopping power
are symmetric about the sign of the charge. That is, an attractive interaction would have the 
same stopping power as a repulsive one does. It is the Coulomb potential specifically, 
which causes this symmetry as the Rutherford scattering cross section depends only on the square of the charges of the interacting particles. The potential of mean force is not 
Coulombic and is screened, thus this symmetry does not exist and results in the Barkas 
effect~\cite{BarkasPRL1963}. This means that the collision rate of an attractive interaction differs from that of a repulsive interaction. This has been studied 
before in classical strongly coupled plasmas~\cite{ShafferPOP2019} as well as strongly 
magnetized plasmas~\cite{JosePOP2022}. Since this work uses the potential 
of mean force, the Barkas effect is expected
to be seen. This is an effect that the analytic and linear response theories do 
not capture as they either assume the force to be Coulombic or have a term 
proportional to the square of the charge, which is the manifestation of this 
symmetry.

\section{Conclusion}

A model is presented for the stopping of ions in warm dense matter based on a combination of 
the BUU equation, which is a semi-classical 
generalization of the classical Boltzmann equation, and the mean force theory, which extends the model to strongly correlated conditions. This work serves as a 
generalization of Ref.~\cite{RightleyPRE2021} to arbitrary ion speeds. The model 
is compared to known theories of stopping power and shown to agree with state of 
the art TDDFT molecular dynamics simulations~\cite{WhitePRB2018}. 

The result is a computationally efficient model that can be used to create tables 
of stopping powers for conditions ranging from weakly coupled classical plasmas to the warm dense matter regime. These could be used as inputs 
to hydrodynamic simulations relevant to ICF to improve the 
predictive power of the simulations. Additionally, the agreement with 
TDDFT in the tail of the stopping power 
curve implies that this work can be used to model experimental 
studies on stopping power, which tend to operate in the high speed tail.

The model was shown to capture the transition between between classical
plasma physics to degenerate plasma physics qualitatively. This transition is 
described by the point $\Theta =1$, where most of the electron scattering occurs at  
energies comparable with the Fermi energy instead of the average thermal energy. 
At this point, the ion-electron collision rate is weakly dependent on temperature, yielding a stopping power curve which does not appreciably change 
as $\Theta$ is further decreased. Qualitatively, this behavior is similar to that 
observed in solid state theories, which are often based on an ideal Fermi gas approximation. 
%which may be surprising as the BUU equation is primarily based on gas kinetic theory concepts.

%Additionally, we commented that we do expect to see the Barkas effect \cite{BarkasPRL1963},
%as a result of using a screened potential instead of the coulomb potential.

Overall, the result is a natural extension of previous works in mean force kinetic 
theory \cite{RightleyPRE2021, BaalrudPOP2019, BaalrudPRL2013} extending
it further into the degenerate regime. Extensions would involve providing 
a systematic expansion of the BUU collision operator to provide a framework for
calculating all transport quantities. Additionally, the effect of collisions between
electrons and inelastic effects were neglected, but can be important for 
more complicated materials and electronic structure.

\begin{acknowledgments}
The authors thank Dr.~Louis Jose and Ryan Park for helpful 
conversations and well as Dr.~Charles Starrett for input data.
This work is funded by the U.S.~Department of Energy NNSA Center of Excellence 
under cooperative agreement number DE-NA0004146 and by 
the Department of Energy [National Nuclear Security Administration] University 
of Rochester “National Inertial Confinement Fusion Program” under award No. 
DE-NA0004144.
%\textcolor{red}{Need acknowledgment for Shane?}
\end{acknowledgments}

\appendix
\section{\label{app:scattering}Quantum Scattering}
The scheme used for quantum scattering calculations is based in the partial 
wave expansion~\cite{Sakurai}, where the scattering cross section is expanded in spherical 
harmonics to give 
\begin{equation}
    \odv{\sigma}{\Omega} = \vab{\sum_{l=0}^\infty (2l+1)\ab(\frac{e^{i\delta_l}\sin\delta_l}{k}) P_l\ab(\cos\theta)}^2,
    \label{eqn:quantum_scattering}
\end{equation}
where $l$ is the angular momentum quantum number, $\delta_l$ is the phase shift,
$\theta$ is the polar scattering angle, $k$ is the magnitude of scattering k-vector defined 
using the energy of the scattering event $E=\hbar^2 k^2/(2 m_e)$, and 
$P_l(x)$ is a Legendre polymonial. Usually, $\delta_l$ is determined 
by solving the radial Schrodinger equation, which can be prone to errors at large
energies. Instead the variable phase method \cite{Calogero1967} is used. This involves solving a non-linear first order differential equation for the 
scattering phase shift (given in atomic units $\hbar = m_e = 1$) by 
\begin{equation}
    \odv{\delta_l}{r}\ab(r) = -\frac{2}{k}V(r)\ab(\hat{j}_l(kr)\cos\delta_l(r)-\hat{n}_l(kr)\sin\delta_l(r))^2,
    \label{eqn:phase_eqn}
\end{equation}
where $\delta_l(r)$ is the phase accumulated by a scattered wave function a radius $r$ away from the scattering center, $k$ 
is the magnitude of the scattering k-vector, and $\hat{j}_l(kr)$ and $\hat{n}_l(kr)$ are Riccati-Bessel
functions defined as $\hat{j}_l(kr) = krj_l(kr)$ and $\hat{n}_l =krn_l(kr)$ where 
$j_l(kr)$ and $n_l(kr)$ are the usual spherical Bessel functions. This equation
has the initial condition that $\delta_l(0) = 0$ and must be integrated to a 
distance that the potential $V(r)$ is sufficiently small so that the system can 
be considered to be in its asymptotic state.

The phase equation Eq.~(\ref{eqn:phase_eqn}) can be a stiff differential equation,
so special solvers must be used. One such solver which has proven robust in this 
problem is LSODA \cite{LSODA}. At higher temperatures, when hundreds of partial
waves are required for convergence, the scheme is changed from solving Eq.~(\ref{eqn:phase_eqn})
to using a Born approximation for the phase shift given by \cite{LinPPFC2023}
\begin{equation}
    \delta_l = -\frac{2}{k}\int_0^\infty \odif{r} V(r) \ab(\hat{j}_l(kr))^2.
    \label{eqn:shift_born}
\end{equation}
The switch between Eq.~(\ref{eqn:phase_eqn}) and Eq.~(\ref{eqn:shift_born}) is 
explained in Ref.~\cite{LinPPFC2023}, and involves predicting when the scattered 
wave functions will have non trivial features the Born-type approximation cannot 
capture. 

At even larger energies, when $\max\ab(rV(r)) < \gamma \hbar^2 k^2/(2m_e)$,
the full Born cross section is used for the entire 
calculation \cite{Sakurai}
\begin{equation}
    \odv{\sigma}{\Omega} = \vab{-\frac{2}{q}\int_0^\infty \odif{r} rV(r) \sin\ab(qr)}^2
    \label{eqn:born}
\end{equation}
where $q = 2k \sin(\theta/2)$, and $\gamma$ is a constant usually set to 0.07. 

Finally, often the differential cross section can be separated from all the other 
integrals and can be put into the momentum transfer cross section defined as 
\begin{equation}
    \sigma_p = \int \odif{\Omega} \odv{\sigma}{\Omega} \ab(1-\cos\theta).
\end{equation}
This can be expanded with Legendre polynomials and reduced in 
the partial wave expansion to 
\begin{equation}
    \sigma_p = \frac{4\pi}{k^2}\sum_{l=0}^\infty (l+1) \sin^2\ab(\delta_{l+1} - \delta_l)
\end{equation}
where $k$ is the k-vector of a collision, $l$ is the angular momentum quantum number 
and $\delta_l$ is the phase shift for a scattered wave with angular momentum $l$.

\section{\label{app:ref}Reference Frame for Numerical Calculations}

When deriving Eq.~(\ref{eqn:final}) from Eq.~(\ref{eqn:pre_ref}), one must choose 
a reference frame for numerical integration. This is fairly complicated as there
are three vectors which must be explicitly written out, and none of them share 
a convenient form. The three vectors are the pre-collision relative velocity 
$\bm{u}$, the change in the relative velocity over the collision $\bm{\Delta u}$,
and the ion velocity in the frame of reference of electrons $\bm{\Delta V}$. Since the only free vector here is the 
ion velocity, this is chosen as the z-axis for the integration. The pre-collision 
relative velocity vector in spherical coordinates is $\bm{u} = u\ab[ \sin\theta^\prime \cos\phi^\prime \hat{x} + 
\sin\theta^\prime \sin\phi^\prime \hat{y} + \cos\theta^\prime \hat{z}]$
where $u = \vab{\bm{u}}$. The change in the relative velocity vector is then the 
complicated one, as it is related to $\bm{u}$ by the scattering angles $\theta$ and 
$\phi$. It is given by 
\begin{widetext}
\begin{align}
    \bm{\Delta u} = u \Big[ &\cos\phi^\prime \ab(\sin\theta\cos\theta^\prime \cos\ab(\phi-\phi^\prime) -\sin\theta\tan\phi^\prime\sin(\phi-\phi^\prime) + (\cos\theta -1) \sin\theta^\prime) \hat{x} \nonumber \\
                            &+\sin\phi^\prime \ab(\sin\theta\cos\theta^\prime \cos\ab(\phi-\phi^\prime) +\sin\theta\cot\phi^\prime\sin(\phi-\phi^\prime) + (\cos\theta -1) \sin\theta^\prime) \hat{y} \nonumber \\
                            &+\ab((\cos\theta-1)\cos\theta^\prime - \sin\theta^\prime\sin\theta\cos(\phi-\phi^\prime))\hat{z} \Big].
    \label{eqn:delta_u}
\end{align}
\end{widetext}
The utility in defining $\tilde{\phi} = \phi-\phi^\prime$ can now be explicitly seen. It can 
also be recognized that the integral over $\phi^\prime$ in Eq.~(\ref{eqn:pre_ref}) will
cause the $x$ and $y$ components of the expression to go to 0, due to the 
$\cos\phi^\prime$ and $\sin\phi^\prime$ terms, indicating that the stopping 
force is along the velocity of the ballistic particle.

\bibliography{refs}
\end{document}